\documentclass[floatfix, aip, jap, superscriptaddress, reprint]{revtex4-1}

\usepackage{graphicx}               
\usepackage{bm}                     
\usepackage{amsfonts,amssymb}       
\usepackage{dcolumn}                
\usepackage{rotating}

\newcommand*\pct{\scalebox{0.9}{\%}}

\begin{document}

\title{
Exploring T-carbon for Energy Applications
}

\author{Guangzhao~Qin}
\affiliation{Department of Mechanical Engineering, University of South Carolina, Columbia, SC 29208, USA}
\author{Kuan-Rong~Hao}
\affiliation{School of Physical Sciences, University of Chinese Academy of Sciences, Beijing 100049, China}
\author{Qing-Bo~Yan}
\email{yan@ucas.ac.cn}
\affiliation{College of Materials Science and Opto-Electronic Technology, University of Chinese Academy of Sciences, Beijing 100049, China}
\author{Ming~Hu}
\email{hu@sc.edu}
\affiliation{Department of Mechanical Engineering, University of South Carolina, Columbia, SC 29208, USA}
\author{Gang~Su}
\email{gsu@ucas.ac.cn}
\affiliation{School of Physical Sciences, University of Chinese Academy of Sciences, Beijing 100049, China}
\affiliation{Kavli Institute for Theoretical Sciences, and CAS Center for Excellence in Topological Quantum Computation, University of Chinese Academy of Sciences, Beijing 100190, China}

\date{\today}


\begin{abstract}
Seeking for next-generation energy sources that are economic, sustainable
(renewable), clean (environment-friendly), and abundant in earth is crucial when
facing the challenges of energy crisis.
There have been numerous studies exploring the possibility of carbon based
materials to be utilized in future energy applications.
In this paper, we introduce T-carbon, which is a theoretically predicted but
recently experimentally synthesized carbon allotrope, as a promising material
for next-generation energy applications.
It is shown that T-carbon can be potentially used in thermoelectrics, hydrogen
storage, lithium ion batteries, \emph{etc}.
The challenges, opportunities, and possible directions for future studies of
energy applications of T-carbon are also addressed.
With the development of more environment-friendly technologies, the promising
applications of T-carbon in energy fields would not only produce scientifically
significant impact in related fields but also lead to a number of industrial and
technical applications.
\end{abstract}

\maketitle

\section{Introduction}
%
%
With the rapid development of human society and global economy, the expense
of resources has increased progressively, especially since the first industrial
revolution\cite{2015_Science_347_1246501_Bonaccorso_Graphene}.
The existing fossil fuels in earth such as coal, oil and natural gas, which were
accumulated in the past billions of years, would be probably exhausted in
hundreds years due to the huge energy demand.
The approaching to resource exhaustion and the accompanying production of
environmentally harmful by-products push us to find possible solutions for
future energy.
This could be remedied in two ways.
One is to promote current utilization efficiency of energy, and to develop
novel technologies to reduce the waste of energy, and to collect waste heat for
reuse.
The other is to find sustainable energy, renewable sources, clean fuels,
\emph{etc}.

It is known that large amount of energy is wasted in factories, home cooking and
vehicle driving because of the low efficiency of energy conversion.
To name a few, the efficiency of engine is about 25-50\pct, where the remaining
part of energy is dispersed to nature in the form of waste heat, which makes
serious environmental pollution and the waste of a lot of resources.
If the waste heat can be recycled, we would improve fundamentally the efficiency
of energy utilization and solve, to some extent, the current energy and
environmental problems.
On the other hand, carbon dioxide and monoxide as well as dust particles such as
PM 2.5 are harmful by-products when consuming fossil fuels, which are
responsible for the global warming due to greenhouse effect and are very harmful
to individual's health in addition to the environmental pollution.
Thus, it is on demands to solve these challenges by seeking for next-generation
energy sources that should be economic, sustainable (renewable), clean
(environment-friendly), and abundant in earth.

\begin{figure*}[tb]
    \centering
    \includegraphics[width=1.00\linewidth]{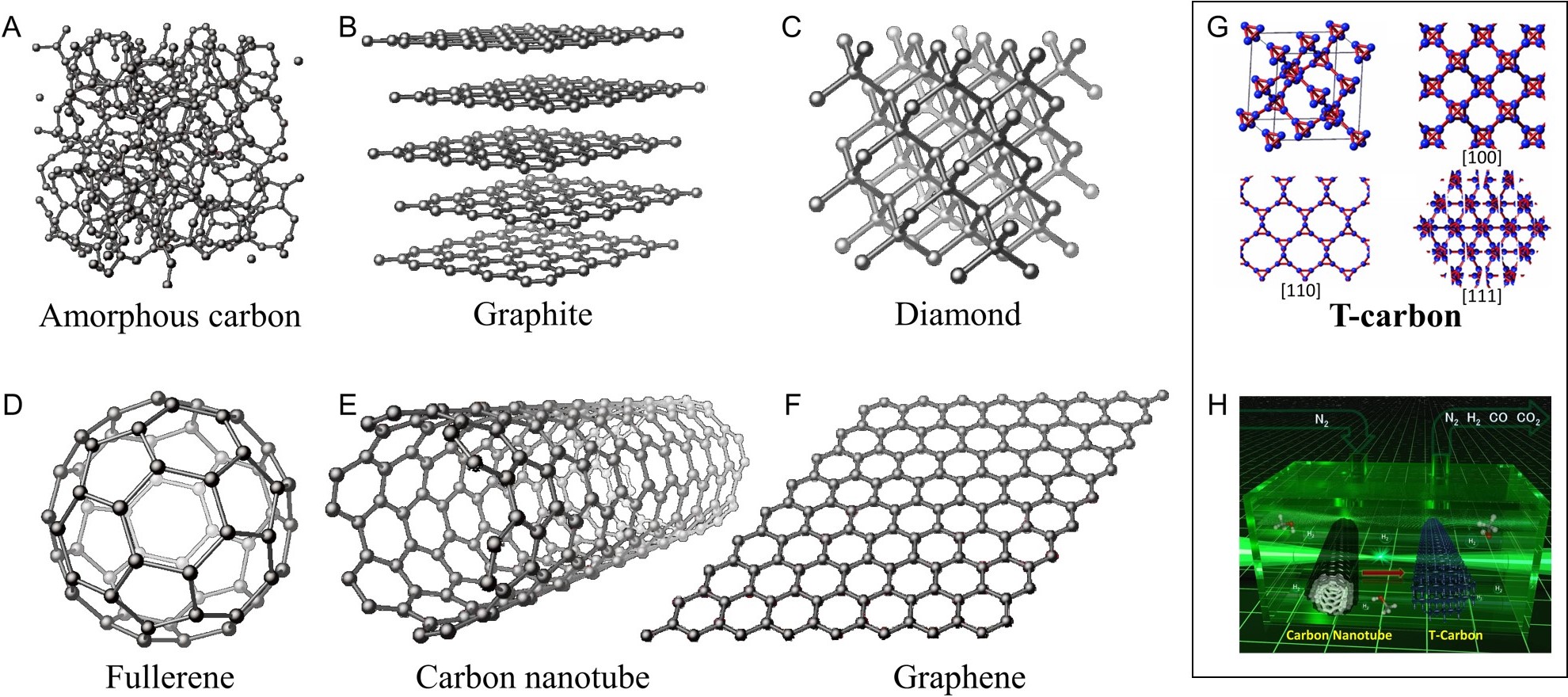}
\caption{\label{fig:carbon}
The position of T-carbon in the carbon family as compared with the common
carbon materials in (A-C) three-dimensional (amorphous carbon, graphite,
diamond), (F) two-dimensional (graphene), (E) one-dimensional (carbon nanotube),
and (D) zero-dimensional (fullerene).
(G) The crystal structure of T-carbon (its space group $Fd\overline{3}m$ is the
same as cubic diamond) is generated by replacing each atom in cubic diamond with
a carbon tetrahedron (C$_4$ unit).
The numbers in [] indicate the crystal direction.
(H) The experimental synthesis layout of T-carbon from a pseudo-topotactic
conversion of multi-walled carbon nanotubes under picosecond pulsed-laser
irradiation in methanol.
Reproduced with permission.\cite{2011_PRL_106_155703_Sheng_TCarbon}
Copyright 2011, American Physical Society.
Reproduction with permission available at
\emph{http://gr.xjtu.edu.cn/web/jinying-zhang/publications}.
}
\end{figure*}

There have been numerous studies exploring the possible utilization of carbon
based materials for next-generation energy applications due to the promising
physical and chemical properties\cite{2018_CAAJ_13_1518_Wang_Recent,
2016_Nanoscale_8_12863_Gao_Electron, 2018_JoIaEC_64_16_Jayaraman_Recent,
2010_NM_9_871_Tour_Green}.
However, large-scale fabrication of carbon materials or carbon based
nanostructures is a formidable challenge\cite{2018_PiEaCS_67_115_Kumar_Recent}.
Lots of efforts have been dedicated to the synthesis processes, such as the
bottom-up approaches from designed carbon
molecules,\cite{2018_ACIE_57_9679_Mori_Carbon}
the pseudo-topotactic conversion of carbon nanotubes by picosecond pulsed-laser
irradiation,\cite{2017_NC_8_683_Zhang_Pseudotopotactic}
\emph{etc}.
Benefited from the progress and emergence of new synthetic technology, the
synthesis of novel carbon materials becomes feasible.
Recently, T-carbon, which is a previously predicted carbon allotrope by
theoretical study, \cite{2011_PRL_106_155703_Sheng_TCarbon}
was experimentally synthesized
(Figure~\ref{fig:carbon}G,H)\cite{2017_NC_8_683_Zhang_Pseudotopotactic}.

\begin{figure*}[tb]
    \centering
    \includegraphics[width=1.00\linewidth]{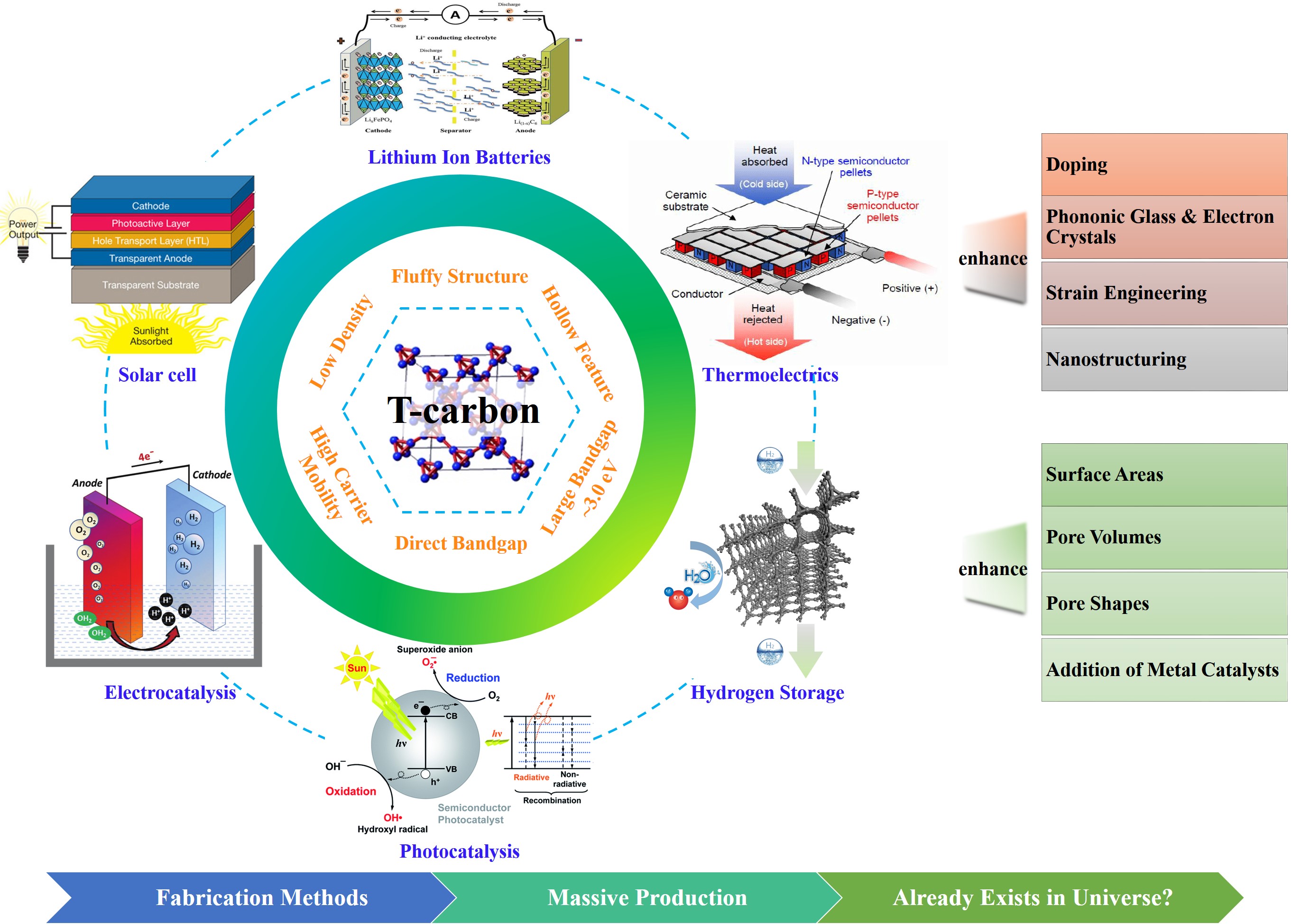}
\caption{\label{fig:overview}
Overview for T-carbon of properties, energy applications, possible enhancement
approaches for thermoelectrics and hydrogen storage, and future development.
Reproduced with permission.
\cite{2014_EES_7_3857_Lee_A, 2016_Nanoscale_8_15033_Joya_Efficient, 2014_MH_1_400_Djurišić_Strategies}
Copyright 2014, 2016, Royal Society of Chemistry.
Reproduction with permission available at
\emph{
    https://www.sigmaaldrich.com/technical-documents/articles/technology-spotlights/plexcore-pv-ink-system.html,
    http://toroccoscoolingandheating.com/thermoelectric-wine-coolers-work/
}
%
%
}
\end{figure*}

\begin{figure*}[tb]
    \centering
    \includegraphics[width=1.00\linewidth]{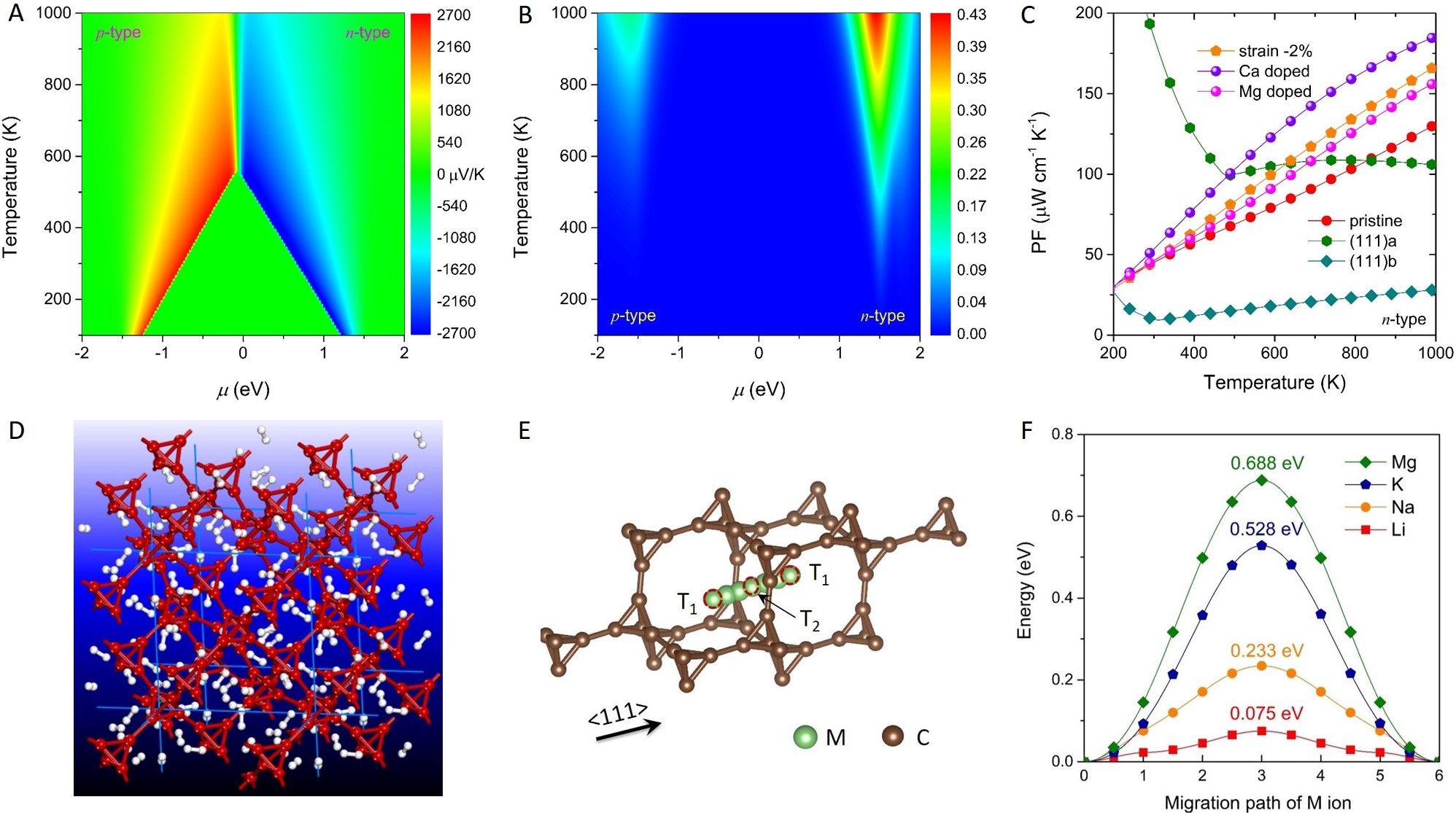}
\caption{\label{fig:energy}
The typical energy applications of T-carbon in (A-C) thermoelectrics, (D)
hydrogen storage, and (E,F) lithium ion batteries (LIB).
(A) Seebeck coefficient (thermopower) and (B) the figure of merit $ZT$ in contour
plot in plane of chemical potential ($\mu$) and temperature.
(C) The modulation of the power factor (PF) of T-carbon with compressive strain of
2\protect\pct\ applied, with calcium (Ca) or magnesium (Mg) doped, or being cut
into two-dimensional structures along the (111) direction.
(D) The hydrogen storage in T-carbon with the capacity of $\sim$7.7\,wt\protect\pct.
(E) The overview of M (= Li/Na/K/Mg) atoms migration in T-carbon.
The minimum migration path is between T$_1$ and the neighboring T$_1$ sites,
which are the center of the vacancy.
T$_2$ indicates the middle point.
(F) The energy profiles of Li, Na, K and Mg atoms diffusing along the minimum
migration path as indicated in (E).
}
\end{figure*}

Herein, we would like to introduce T-carbon and discuss its promising
applications for next-generation energy technologies
(Figure~\ref{fig:overview} and Figure~\ref{fig:carbon}).
It is shown that T-carbon can be potentially used in thermoelectrics, hydrogen
storage, lithium ion batteries (Figure~\ref{fig:energy}), \emph{etc}.
The challenges, opportunities, and possible directions for future studies of
energy applications of T-carbon are also addressed.

\section{T-carbon}

Carbon is one of the most abundant elements on earth and one of the most
important elements for life, which is contained in majority of molecules.
Carbon atoms possess unique ability to form bonds with other carbon atoms and
nonmetallic elements in diverse hybridization states ($sp$, $sp^2$,
$sp^3$)\cite{2017_MT_20_592_Borchardt_Toward, 2018_JoIaEC_64_16_Jayaraman_Recent}.
Countless carbon-based organic compounds in the form of a wide range of
structures from small molecules to long chains are generated, which have great
diversity in the chemical and biological properties and thus result in the
present colorful world\cite{2018_CAAJ_13_1518_Wang_Recent}.
Carbon mainly exists as three natural allotropes, namely graphite, diamond, and
amorphous carbon
(Figure~\ref{fig:carbon}A,B,C)\cite{2015_CR_115_4744_Georgakilas_Broad}.
Despite the same and exclusive component of carbon atoms, their properties are
drastically different from each other, which is a strong hint of the diversity
in the properties of carbon materials with different structures and orbital
hybridizations.
Over the last several decades, several new carbon allotropes have been
synthesized with novel properties and potential applications in technology.
The three most typical examples include
zero-dimensional (0D) fullerenes discovered in 1985 (Figure~\ref{fig:carbon}D),
one-dimensional (1D) carbon nanotubes identified in 1991 (Figure~\ref{fig:carbon}E),
and two-dimensional (2D) graphene isolation in 2004
(Figure~\ref{fig:carbon}F)\cite{2010_NM_9_868_Hirsch_The}.
The fantastic properties of these carbon allotropes and their highly expected
engineering probabilities have attracted intensive attention from both academia
and industry \cite{2010_NM_9_871_Tour_Green}.

Beyond that, a number of three-dimensional (3D) carbon allotropes have also been
predicted theoretically, including
M-carbon,\cite{2009_PRL_102_175506_Li_Superhard}
bct-C$_4$,\cite{2010_PRL_104_125504_Umemoto_BodyCentered}
BCO-C$_{16}$,\cite{2016_PRL_116_195501_Wang_BodyCentered} \emph{etc}, where
T-carbon predicted in 2011 \cite{2011_PRL_106_155703_Sheng_TCarbon}
is the most impactful form (Figure~\ref{fig:carbon}G).
T-carbon can be simply derived by substituting each carbon atom in cubic diamond
with a C$_4$ unit of carbon tetrahedron (Figure~\ref{fig:carbon}C,G), and it is where
the name of `T-carbon' comes from.
The space group of T-carbon is $Fd\overline{3}m$, the same as cubic diamond.
There exist two tetrahedrons (eight carbon atoms in total) in a unit cell.
Such geometric configuration of carbon atoms that forms 3D T-carbon is
thermodynamically stable, which is confirmed by the non-imaginary frequency of
the phonon dispersion in previous study\cite{2011_PRL_106_155703_Sheng_TCarbon}.
The lattice constant of the fully optimized T-carbon is about 7.52\,\AA, which
is more than two times that of diamond (3.566\,\AA).
As compared to the bond length in diamond (1.544\,\AA), the bonds in T-carbon
possess two types with the bond length being 1.502 and 1.417\,\AA\ for
intratetrahedron and intertetrahedron,
respectively\cite{2017_Carbon_121_154_Esser_Bonding}.
Besides, different from the bond angle in diamond (109.5$^\circ$), the bond angle
in T-carbon are 60 and 144.74$^\circ$ for the bonds in tetrahedron and two
inequivalent bonds, respectively, implying the existence of strain.
Because of the large interspaces between atoms in T-carbon, the density is
1.50\,g/cm$^3$, which is much smaller compared to that of diamond, graphite,
M-carbon, and bct-C$_4$\cite{2011_PRL_106_155703_Sheng_TCarbon}.
In addition to the low density, the Vickers hardness of T-carbon is calculated
to be 61.1\,GPa, which is around 1/3 softer than diamond
(96\,GPa)\cite{2011_PRB_84_121405_Chen_Hardness, 2011_PRL_106_155703_Sheng_TCarbon}.
The low density with large interspaces between atoms and the soft nature would
promise broad applications of T-carbon.

T-carbon has recently been synthesized in experiment (Figure~\ref{fig:carbon}H)
from a pseudo-topotactic conversion of multi-walled carbon nanotube (MWCNTs)
suspended in methanol under picosecond pulsed-laser
irradiation \cite{2017_NC_8_683_Zhang_Pseudotopotactic}.
Firstly, MWCNTs (length: $~\sim$100-200\,nm; diameter: $\sim$10-20\,nm) are
prepared by chemical vapor deposition (CVD) and subsequent processing.
After dispersed in absolute methanol, the suspension containing individualized
MWCNTs are then transferred to the self-designed process with laser irradiation,
which is stirred with a magnetic stirring bar and kept under the nitrogen
atmosphere.
In the fast and far-from-equilibrium process, the metastable structure is
captured with the successfully transition from $sp^2$ to $sp^3$ chemical bonds
(Figure~\ref{fig:carbon}H) and the suspension becomes transparent after the laser
reaction.
Hollow carbon nanotubes are transformed into solid carbon nanorods, where the
connections between carbon atoms are exactly the same as the theoretical
predicted T-carbon, demonstrating the synthesis of this kind of structure.
During the transformation process, the time scale of energy transfer from the
laser to the MWCNTs and the subsequent ultrafast quenching play a key role in
the formation and stabilization of T-carbon.
The cubic crystal system of the generated T-carbon NWs is confirmed by the fast
Fourier transform (FFT) pattern at different tilting angles from the
high-resolution transmission electron microscopy (HRTEM) image.
The successful synthesis of T-carbon makes it joining in the carbon family as
another achievable 3D carbon allotrope in addition to graphite, diamond, and
amorphous carbon (Figure~\ref{fig:carbon}).

T-carbon's proposal and then experimental realization is a breakthrough in
carbon science\cite{2016_Materials_9_484_Xing_A, 2016_JPCM_28_475402_Wang_C20}.
Compared with other allotropes of carbon, T-carbon has many unique and
intriguing properties (Figure~\ref{fig:overview}), suggesting that it could have
a wide variety of potential applications in photocatalysis, solar cells,
adsorption, energy storage, supercpacitors, aerospace materials, electronic
devices, \emph{etc}.
For example, T-carbon is predicted to be a semiconductor with direct band gap of
$\sim$3.0\,eV at $\Gamma$-point (GGA: 2.25\,eV; HSE06: 2.273\,eV; B3LYP:
2.968\,eV)\cite{2019_nCM_5_9_Sun_A, 2011_PRL_106_155703_Sheng_TCarbon}.
The orbitals in T-carbon hybridize with each other and form anisotropic $sp^3$
hybridized bonds.
As for the two types of bonds in T-carbon, the charge density is found to be
much larger for the intertetrahedron bonds compared to the intratetrahedron
bonds, indicating relatively stronger intertetrahedron bond strength with
more accumulated electrons.
The bond strength is consistent with the bond length, which stabilize the
structure by balancing the strain from the carbon tetrahedron cage.
Moreover, the band gap could be effectively adjusted by doping elements or
strain engineering to be suitable for photocatalysis and
solar cells\cite{2019_Optik_180_125_Alborznia_Pressure, 2019_CP_518_69_Ren_Efficient, 2019_nCM_5_9_Sun_A}.
Particularly, the band gap can be tuned in the range of 1.62-3.63\,eV with group
IVA single-atom substitution, where the doped structures hold the
stability\cite{2019_CP_518_69_Ren_Efficient}.
Quite recently, it was shown that T-carbon nanowire exhibits better ductility
and larger failure strain than other carbon materials such as diamond and
diamond-like carbon \cite{2018_Carbon_138_357_Bai_Mechanical}.
It was also reported that the transport properties of T-carbon can be
effectively modulated by imposing strain\cite{2019_nCM_5_9_Sun_A}.
With the specific characteristics of the electronic band structures such as the
potential efficient electron transport,\cite{2019_nCM_5_9_Sun_A} T-carbon has
the potential to be used as thermoelectric materials for energy recovery and
conversion, \cite{2017_PRB_95_085207_Yue_Thermal} especially after doping or
strain engineering.
Besides, owing to its `fluffy' crystal structure, T-carbon is also possible for
the storage of hydrogen, lithium, and other small molecules for energy purposes.
In the following, the specific applications of T-carbon in thermoelectrics,
hydrogen storage, and lithium ion batteries will be discussed in details to
illustrate its potential applications in future energy fields.

\section{Thermoelectrics}

In the sense of `turning waste into treasure', the thermoelectric
power generation has received extensive attention in recent years due to the low
cost of operation.
By achieving the output voltage through temperature gradient based on the
Seebeck effect, the thermoelectrics shows a strong capability of firsthand
solid-state conversion to electrical power from thermal energy, especially from
the reuse of waste heat \cite{2012_Nature_489_414_Biswas_Highperformance},
thereby revealing its valued applications in reusing resources and being helpful
for the crisis of environment and the save of energy.
%
Generally, the thermoelectric efficiency and performance can be characterized by
a dimensionless figure of merit $ZT = S^2\sigma T/\kappa$,
\cite{2014_SR_4_6946_Guangzhao_Hingelike}
where $S$, $\sigma$, $T$ and $\kappa$ represent the thermopower (Seebeck
coefficient), electrical conductivity, absolute temperature and total thermal
conductivity, respectively.
To approach the Carnot coefficient, a high energy generation efficiency is
necessary, which corresponds to a large $ZT$.
The continuous improvement of thermoelectric performance and the strive to
increase the power output under the same heat source are the key focus in the
thermoelectric technology, which demands the in-depth study of thermoelectric
conversion materials and the development of new materials.

Based on electronic structures and previously studied thermal transport
properties of T-carbon \cite{2017_PRB_95_085207_Yue_Thermal}, we examined the
thermoelectric performance of T-carbon by combining the first-principles
calculations with the semi-classical Boltzmann transport
theory\cite{2014_SR_4_6946_Guangzhao_Hingelike, 2006_CPC_175_67_Georg_BoltzTraP}.
The thermopower of T-carbon shown in Figure~\ref{fig:energy}A ($\sim$2000\,$\mu$V/K)
is comparable with or even larger than some excellent thermoelectric materials,
such as SnSe ($\sim$550\,$\mu$V/K) \cite{2016_EES_9_3044_Zhao_SnSe} that
was reported to have an unprecedented high $ZT$ value.
The huge thermopower of T-carbon indicates its strong potential capable of
serving as a thermoelectric material for energy recovery and conversion.

The overall view of evaluated $ZT$ value of T-carbon shows that it is a
high-temperature $n$-type thermoelectric material (Figure~\ref{fig:energy}B).
However, the thermoelectric performance of T-carbon is not good enough compared
with other existing thermoelectric materials \cite{2012_Nature_489_414_Biswas_Highperformance}.
For instance, SnSe possesses a $ZT$ value of 2.6 at 930\,K along a specific lattice
direction \cite{2016_EES_9_3044_Zhao_SnSe}.
The reasons could lie in two aspects.
Firstly, the electronic energy band gap of T-carbon is relatively large, and the
conduction band minimum (CBM) and valence band maximum (VBM) are relatively
flat, which may lead to a large effective mass of the carriers and lower the
electrical conductivity.
Secondly, the thermal conductivity of T-carbon at room temperature is 33\,W/mK,
\cite{2017_PRB_95_085207_Yue_Thermal} which is much higher for thermoelectric
applications in contrast to that of SnSe (0.46-0.68\,W/mK)
\cite{2016_EES_9_3044_Zhao_SnSe}.

Nevertheless, there exists a large room for further improving the
thermoelectric performance of T-carbon in view of its excellent thermopower
through, \emph{e.g.}\ applying strain,\cite{2014_SR_4_6946_Guangzhao_Hingelike}
doping proper elements,\cite{2016_SR_6_26774_Gharsallah_Giant} or cutting into
low-dimensional structures to modify transport
properties\cite{2016_Nanoscale_8_11306_Qin_Diverse}.
We then examined possible approaches for the improvement of the thermoelectric
performance of T-carbon.
As shown in Figure~\ref{fig:energy}C, by either applying compressive strain of
2\pct\ or doping calcium (Ca) and magnesium (Mg) atoms into the fluffy structure
of T-carbon, the power factor can be effectively improved.
We focus on the $n$-type doping, since T-carbon is a $n$-type thermoelectric
material as discussed above.
The reason for the doping enhanced power factor lies in two aspects.
On the one hand, with Ca/Mg atoms doped, the characteristics of conduction band
is kept, promising a large thermopower after doping.
On the other hand, the electronic band gap transits from direct to indirect and
is largely decreased, leading to a large electrical conductivity.
Note that the thermal conductivity commonly decreases with foreigner atoms
doped, thus the thermoelectric performance of T-carbon would be largely improved
owing to the simultaneously improved electrical transport properties and
reduced thermal transport properties.

Considering the huge computational costs, we could estimate the thermoelectric
performance based on the estimated thermal conductivity with atoms doped.
Assuming a one order of magnitude decrease of the thermal conductivity with
Ca/Mg atoms doped, the $ZT$ value of T-carbon is estimated to be largely enhanced
with a two-fold increase.
Moreover, by cutting T-carbon into two-dimensional structures along the (111)
direction, the power factor can also be improved, especially at low
temperatures (Figure~\ref{fig:energy}C).
The large power factor at low temperature makes T-carbon in nanoscale transit
from the high-temperature thermoelectric material to a low-temperature
thermoelectric material, suggesting its wider applications for energy
conversion, such as the waste heat recovery under ambient conditions.

\section{Hydrogen Storage}


Seeking for sustainable, renewable, and clean fuels is on demand, in particular
we are facing with the challenges of energy crisis and climate change.
Since 1970s, hydrogen has been thought as one of the most promising alternatives
to fossil fuels due to the cleanliness of its combustion.
Water (H$_2$O) is the only by-product for hydrogen combustion, which has great
advantages compared with the combustion of fossil fuels producing greenhouse gas
and harmful pollutants.
Moreover, hydrogen is lightweight, providing a higher energy density and making
the hydrogen-powered engines more efficient than the internal combustion
engines.
The reason hampering the popularization of hydrogen economy lies in that it is
difficult to store large amounts of hydrogen safely, densely, rapidly, and then
access easily.
Lots of efforts have been dedicated to discovering new materials as next
generation hydrogen storage materials, including the extremely porous
metal-organic framework (MOF) compounds \cite{2012_CR_112_782_Suh_Hydrogen}.

Benefiting from the high surface area and lightweight of carbon materials, there
have been many efforts in designing novel porous carbon materials for hydrogen
storage applications\cite{2016_Nanoscale_8_12863_Gao_Electron,
2017_MT_20_629_Xu_Design, 2017_MT_20_592_Borchardt_Toward,
2012_TJoPCC_116_25015_Srinivasu_Electronic}.
Since T-carbon itself is a fluffy carbon material, there are large interspaces
between atoms compared with other forms of carbon materials, which could make it
potentially useful for hydrogen storage (Figure~\ref{fig:overview} and
Figure~\ref{fig:energy}D).
In fact, T-carbon possesses a low density ($\sim$1.50\,g/cm$^3$) as mentioned
above, which is approximate 2/3 of graphite and 1/2 of
diamond\cite{2011_PRL_106_155703_Sheng_TCarbon}.
By absorbing hydrogen into the fluffy structure of T-carbon, the hydrogen
storage value can be estimated based on the number of adsorbed hydrogen
molecules (H$_2$), which is 16 in maximum for one unit cell in a stable
structure.
The adsorption energy for hydrogen in T-carbon is defined as\cite{2012_TJoPCC_116_25015_Srinivasu_Electronic}
\begin{equation}
E_\textrm{adsorption} = [E_\textrm{T-carbon} + nE_\mathrm{H_2} - E_\textrm{total}] / n \ ,
\end{equation}
where $E_\textrm{T-carbon}$ is the total energy of T-carbon, $E_\mathrm{H_2}$ is
the total energy of hydrogen molecules, $n$ is the number of hydrogen molecules,
and $E_\textrm{total}$ is the total energy of T-carbon with hydrogen absorbed.
$E_\textrm{adsorption}$ is 0.173 and -0.216\,eV for 8 and 16 H$_2$ absorbed,
respectively.
Considering the strong C-C bonding in T-carbon, the system is stable despite the
weak repulsive interactions among the absorbed H$_2$ like in fullerenes with
hydrogen absorbed.
With this condition, the hydrogen storage capacity of T-carbon is estimated to be
$\sim$7.7\,wt\pct, which makes it quite competitive for the high-capacity
hydrogen storage \cite{2011_PRL_106_155703_Sheng_TCarbon}.

\section{Lithium Ion Batteries}


Rechargeable energy storage devices such as lithium ion batteries (LIB) are
playing a critical role as portable power sources in electronic devices,
biomedicine, aviation space and electric vehicles\cite{2015_MT_18_252_Nitta_Liion}.
Various carbon based materials have been widely used in LIB, among which
graphite is the most commonly used anode material.
Due to its layered structure with high specific surface area and large
interlayer space to accommodate lithium atoms, graphite has a high
specific energy capacity (372\,mAhg$^{-1}$)\cite{2010_NL_10_2838_Uthaisar_Edge}.

As a new member of carbon materials family, T-carbon could be also a promising
electrode material for LIB and other rechargeable energy storage devices due to
its fluffy structure.
The possibilities of T-carbon acting as an electrode material for alkali metals
(Li, Na, K) and alkaline earth metal (Mg) ion batteries were investigated based
on first-principles calculations.
The specific capacity of metal atoms is defined as
$C=nF/M_{C_X}$,\cite{2015_MT_18_252_Nitta_Liion} where $n$ represents the number
of electrons involved in the electrochemical process ($n$=1 for Li, Na, K; $n$=2
for Mg), $F$ is the Faraday constant with a value of 26.801\,Ahmol$^{-1}$,
$M_{C_X}$ is the mass of $C_X$ ($C$ means carbon atoms, $X$ means the number of
carbon atoms, which is 8 for T-carbon).
Our results reveal that T-carbon is a good anode material for LIB with the
specific energy capacity of 588\,mAhg$^{-1}$, which is 58\pct\ higher than that
of graphite (372\,mAhg$^{-1}$)\cite{2010_NL_10_2838_Uthaisar_Edge}.
The corresponding formula is Li$_2$C$_8$, indicating that two Li ions can be
intercalated in each T-carbon unit cell.
The results for Na, K, and Mg are also similar, except that the specific energy
capacity for Mg is 1176\,mAhg$^{-1}$ due to its doubled valence electrons.

As shown in Figure~\ref{fig:energy}E, the most stable adsorption site of metallic
ions was calculated to be the center of the vacancy of T-carbon, which is marked
as T$_1$ site.
The migration process of M (= Li/Na/K/Mg) ions in T-carbon was simulated by
means of the climbing image nudged elastic band method
(CI-NEB)\cite{2000_JCP_113_9901_HenkelmanGraeme_A} in a $2\times 2\times 2$ supercell.
As indicated in Figure~\ref{fig:energy}E, the minimum migration path is between
the neighboring T$_1$ sites.
The middle point (T$_2$ site) corresponds to the saddle point on the potential
energy surface.
Figure~\ref{fig:energy}F shows the energy evolution for the migration process,
where the migration barriers are 0.075, 0.233, 0.528, and 0.688\,eV for Li, Na,
K, and Mg, respectively.
The lowest barrier for ion moving from T$_1$ to neighboring T$_1$ sites in
T-carbon is observed for Li ion.

It should be noted that the Li migration barrier in T-carbon is about 1/4 of
that in graphite (0.327\,eV) \cite{2010_NL_10_2838_Uthaisar_Edge}, which implies
that the diffusion constant of Li ion in T-carbon should be $1.7\times 10^4$
times larger than that of Li ion in graphite following the Arrhenius law
($D\sim \exp (-E/k_B T)$, where $E$, $k_B$, and $T$ are barrier energy,
Boltzmann constants, and temperature,
respectively)\cite{2018_PCCP_20_9865_Hao_Lithium}.
Thus, T-carbon should be a very good material for the diffusion of Li ions,
revealing that it might be quite useful for the ultrafast charge and discharge
of future rechargeable energy storage devices.

\section{Opportunities and Challenges}

Both opportunities and challenges exist for the applications of T-carbon in
next-generation energy technologies (Figure~\ref{fig:overview}).
To achieve better thermoelectric performance of T-carbon, doping other elements
can be performed by following what researchers did previously for clathrates and
skutterudites,\cite{2003_IMR_48_45_Chen_Recent} which are hot thermoelectric
materials with hollow cage-like structures.
In terms of the fluffy structure with low density and hollow feature
(Figure~\ref{fig:overview}), the characteristics of `phononic glass \& electron
crystals' could be realized in T-carbon by filling the holes with different
kinds of atoms and different filling rates, thus reducing its thermal
conductivity and simultaneously improving its electrical transport properties,
which would make T-carbon a better thermoelectric material.
In addition to the filling doping, one can also introduce nanotwin structures
into T-carbon, which could have the similar effect \cite{2017_Nanoscale_9_9987_Zhou_Decouple}.
Based on previous works, the possible interstitial atoms to improve the
thermoelectric performance of T-carbon could be lanthanides, alkali metals,
alkaline earth metals, and rare earth metals.
Other approaches in addition to
applying external fields\cite{2016_Nanoscale_8_11306_Qin_Diverse}
and bond nanodesigning,\cite{2018_NE_50_425_Guangzhao_Lonepair, 2016_PRB_94_165445_Qin_Resonant}
such as strain engineering and nanostructuring, would also be possible
(Figure~\ref{fig:energy}C).
Further detailed and comprehensive studies for examining possible improvement of
the thermoelectric performance, especially from the experimental aspects, are
expected to come in future.

In addition, with the potentially high capacity of hydrogen storage in T-carbon,
the effects of different surface areas, pore volumes, and pore shapes on
hydrogen storage parameters should be examined.
New methods to enhance the storage capacity are necessary, such as the addition
of metal catalysts, which has been previously reported for considerably improving
the capacity of hydrogen storage.
With the fluffy structure, it is also possible for T-carbon to be used for
storing or filtering other small molecules for energy purposes beyond
the applications in LIB.
Beyond the applications of T-carbon in thermoelectrics, hydrogen storage, and
lithium ion batteries as discussed above, T-carbon, especially T-carbon based
heterostructures,\cite{2018_JoIaEC_64_16_Jayaraman_Recent,
2018_ACIE_57_9679_Mori_Carbon, 2018_Carbon_137_266_Ram_Tetrahexcarbon}
could have a wide variety of potential applications in more energy fields
(Figure~\ref{fig:overview}), such as electrochemical, photocatalysis, solar
cells, adsorption, energy storage, supercpacitors, aerospace, electronics,
\emph{etc}, which are worth to be further investigated in future.


To further explore potential applications of T-carbon in energy fields, much
effort on fabrication methods and massive production of T-carbon should be
greatly paid (Figure~\ref{fig:overview})\cite{2018_PiEaCS_67_115_Kumar_Recent}.
Apart from the synthesizing method reported in Ref.~\cite{2017_NC_8_683_Zhang_Pseudotopotactic},
the plasma enhanced chemical vapor deposition at a proper environmental pressure
is also a possible route to generate T-carbon.
Particularly, T-carbon is found to be more stable and more easily formed at
negative pressure circumstances since it possesses a relatively smaller enthalpy
than diamond beyond negative 22.5\,GPa.
In addition, T-carbon could be grown from seed microparticles in a chemical
vapor transport process, where the seed quality and distribution should be
optimized to obtain high-quality T-carbon samples.
With the development of more environment-friendly technologies, the potential
applications of T-carbon in energy fields would not only produce scientifically
significant impact in related fields but also lead to a number of industrial and
technical applications.

Beyond the applications in energy fields, T-carbon may also contribute to
solving the carbon crisis in interstellar dust\cite{1997_TAJ_484_779_Dwek_Can},
which remains an unsolved question for decades.
Observations find that the abundance of carbon in the interstellar medium is
only $\sim$60\pct\ of its solar value, leading to the difficulty in explaining
the interstellar extinction curve with the traditional interstellar dust
models\cite{1997_TAJ_484_779_Dwek_Can}.
Due to the fluffy structure of T-carbon, its density is approximately 2/3 of
graphite.
Besides, the optical absorption of T-carbon has a sharp peak around 225\,nm,
which is very close to the broad `bump' centered at 217.5\,nm in the
interstellar extinction curve\cite{2017_NC_8_683_Zhang_Pseudotopotactic,
1997_TAJ_484_779_Dwek_Can}.
Moreover, the negative pressure circumstance in the universe is beneficial for
the formation of T-carbon.
Thus, it would be very meaningful to explore whether T-carbon already exists in
universe beyond the artificial synthesis.

\section*{Acknowledgments}
G.\ Q.\ gratefully acknowledges
    Dr.\ Zhenzhen Qin (Zhengzhou University) for literature review and
    Dr.\ Huimin Wang (Nanjing University) for plotting Figure~\ref{fig:overview}.
G.\ Q.\ also thanks
    Dr.\ Xianlei Sheng (Beihang University)
    and Mr.\ Jingyang You (University of Chinese Academy of Sciences)
for their fruitful discussions.
This work was supported in part by the National Key R\&D Program of China
(Grant No.\ 2018YFA0305800), the NSFC (Grant Nos.\ 11834014, 14474279), the
Strategic Priority Research Program of the Chinese Academy of Sciences (Grant
Nos.\ XDB28000000, XDPB08).

\section*{Computation details}
All calculations involved in this paper (Figure~\ref{fig:energy}) were carried
out by means of the first-principles calculations in the framework of density
functional theory (DFT) as implemented in the Vienna \emph{ab-initio} simulation
package (VASP)\cite{1996_PRB_54_11169_uller_Efficient}.
The projector augmented wave (PAW) method\cite{1999_PRB_59_1758_Kresse_From}
were employed for interactions between ion cores and valence electrons.
The electron exchange-correlation interactions were described by the generalized
gradient approximation (GGA) in the form proposed by Perdew-Burke-Ernzerhof
(PBE)\cite{1996_PRL_77_3865_Perdew_Generalized}.
The cutoff energy was set as 1000\,eV for plane-wave expansion of valence
electron wave function.
The structure relaxation considering both atomic positions and lattice vectors
was performed until the total energy was converged to 10$^{-8}$\,eV/atom and the
maximum force on each atom was less than 0.001\,eV/\AA.
The Monkhorst-Pack scheme\cite{1976_PRB_13_5188_Monkhorst_Special} was used to
sample the Brillouin zone (BZ) with a $11\times 11\times 11$ $k$-point mesh.

\bibliography{new_bibliography_add}
\bibliographystyle{nature}

\end{document}